\documentclass{article}
\usepackage{interspeech2012}
\usepackage{amsmath,amssymb,eucal,graphicx}
\usepackage{amsthm}
\usepackage{amsfonts}
\usepackage{float}
\usepackage{mathcomSTEP}

\ninept
\def\reg{{\rm\ooalign{\hfil
     \raise.07ex\hbox{\scriptsize R}\hfil\crcr\mathhexbox20D}}}

\title{Improving the Entropy Estimate of Neuronal Firings of Modeled Cochlear Nucleus Neurons}

\makeatletter
\def\name#1{\gdef\@name{#1\\}}
\makeatother
\name{{\em Andrea Grigorescu$^1$, Marek Rudnicki$^2$}\\
{\em Michael Isik$^2$, Werner Hemmert$^2$, Stefano Rini$^1$}\\
}

\address{$^1$Institute for Communications Engineering,
  $^2$Institute of Medical Engineering   \\
  Technische Universit\"at M\"unchen, Germany \\
{\small \tt  andrea.grigorescu@mytum.de, marek.rudnicki@tum.de,} \\ {\small \tt michael.isik@tum.de, werner.hemmert@tum.de, stefano.rini@tum.de}}

\begin{document}

\maketitle
\begin{abstract}
In this correspondence information theoretical tools are used to investigate the statistical properties of modeled cochlear nucleus globular bushy cell spike trains.
%
The firing patterns are obtained from a simulation software that generates sample spike trains from any auditory input.
Here we analyze for the first time the responses of globular bushy cells to voiced and unvoiced speech sounds.
Classical entropy estimates, such as the direct method, are improved upon by considering a time-varying and 
time-dependent entropy estimate.
%
%
%
%
With this method we investigated the relationship between the predictability of the neuronal response and the frequency content in the auditory signals.
%
The analysis quantifies the temporal precision of the neuronal coding and the memory in the neuronal response. 
\end{abstract}

\section{Introduction}
The auditory system is an ideal model to study coding and processing of information in the neuronal system.
The fidelity of coding is thereby remarkable: the human ear covers a dynamic range larger than 120\,dB, a frequency range from 16\,Hz to 16\,kHz and provides temporal precision in the order of tens of microseconds, which allows us to
localize sound in the horizontal plane with exquisite resolution.
It is clear that these features require delicate preprocessing in the inner ear.
The key-features are narrow-band filtering and nonlinear dynamic compression, which both depend on an active, mechanical feedback amplification process.
Every location in the inner ear has its characteristic frequency (CF), where a receptor cell transduces mechanical vibrations into an electrical potential, which is then transferred at the chemical synapse into all-or-none nerve-action potentials of the auditory nerve.
A single receptor cell is innervated with multiple (up to 40) auditory nerve fibers (ANFs).
ANFs are classified by their spontaneous rates into three groups: low-, medium- and high spontaneous rate fibers.
They have different thresholds and cover different dynamic ranges.
Neuronal processing starts in the first station of the central nervous system, the cochlear nucleus (CN).
It consists of several neuron types that receive direct inputs from ANFs and show various firing properties.
Here we focus on globular bushy cells (GBC) that are one of the principal cells in CN and known for their precise temporal coding (see \cite{Rhode2008Response} for an overview).
They are involved in the sound localization pathway and are among the fastest neurons in our brain.
They lock on low-frequency tones and amplitude modulated signals like speech.
GBCs receive input from multiple ANFs through giant synapses (endbulbs of Held).
They suppress spontaneous activity and enhance temporal precision by coincidence detection.
Compared to ANFs, their responses are much more reliable, which makes them ideal candidates for information calculations.
Information theory provides important insight on the coding performed by the auditory system and is a theoretical tool commonly
used to interpret 
neuronal data.

%
For example, \emph{entropy} is used as a measure of information in a spike train \cite{strong1998entropy}.
%
%
Entropy specifies the number of bits required to represent a sequence of outcomes of a random source.
In other words, it captures the \emph{richness} of the random output: sources with low entropy produce predictable outcomes, 
while sources 
with high entropy are harder to predict.
When evaluating the entropy of a spike train, we are evaluating the predictability of this random sequence over time.
%
This is an indicator of the level of activity of the neuron, but from this measure one cannot infer much about the origin of this variability or the ``information content'' of the spike train.
%
%

Estimating entropy is, in general, a complex task that may require a  large amount of data and one should not embark 
such an endeavor without a good motivation to do so.

Some of the desirable properties of entropy as a measure of predictibility of 
random variables are the following:
\begin{description}
  \item[\bf Entropy does not change under injective transformations]  Any transformation of the data that can be inverted preserves entropy
  \item[\bf Entropy is bounded] Entropy of discrete alphabet sequences is bounded between a maximum and a minimum which provides an absolute measure of predictability. 
  \item[\bf Entropy can be factorized] When computing the entropy of multiple random variables, one can divide the calculation into sub-tasks.
\end{description}

The so-called estimation of the entropy has been often performed using the direct method \cite{strong1998entropy} which is actually an upper bound to the entropy estimate as well as being a time invariant measure.
This estimate is accurate if the neuronal activity is modeled as
coming from a random, memoryless 
source that has no particular relationship with a known input process.
This is clearly not the case in many scenarios in which
one can obtain a better estimate of the entropy by considering the time dependence and the time variability of the neuronal activity.
By considering these two refinements to the direct method we improve the quality of the entropy estimate.
%
%
  In Sec. \ref{sec:model} we introduce the model we utilize to produce neuronal firings and the auditory inputs we consider.
  In Sec. \ref{sec:Entropy} we introduce fundamental properties of the entropy.
An improvement of the direct method to estimate  the entropy is provided in Sec. \ref{sec:Entropy Estimation}.
and in  Sec. \ref{sec:Numerical Results} we analyze the utterance of a vowel and a consonant by using the entropy estimate 
from Sec. \ref{sec:Entropy Estimation}.
Finally, Sec. \ref{sec:Conclusions} concludes the paper.
%
%
%
%
%
%
%
%
%
%
%
%
%
%
%
%
%
%
%
%
\section{Model Description}
\label{sec:model}
We apply an inner ear model from \cite{Zilany2009Phenomenological}, where we have adopted the middle ear to tune the model to achieve human-like hearing thresholds.
The inner ear model generates random spike trains of ANFs, which excite GBCs.
Our GBC model is realized as a single compartment model with Hodgkin-Huxley like ion channels (HPAC, Kht, Klt) described in \cite{Rothman2003Roles}.
We tuned innervation (32 high-, 4 medium- and 4 low-spontaneous rate fibers) and synaptic weights of our model to replicate low spontaneous activity,
high synchronization and entrainment values (see \cite{Rudnicki2012Are}) according to physiological recordings from \cite{Joris1994EnhancementA}.
This model was also able to reproduce experimental results obtained from pure tone stimulations: PSTH, ISIH and receptive field maps.
The model produces spike-trains to auditory inputs with a time resolution of about 20.8\,$\mu$sec, which were re-sampled with a 1\,ms precision in this study.
%

%
%
In the following we exemplarily 
analyze neuronal firing produced by the 
model at a single CF for two utterances from the Isolated Letter Speech Recognition (ISOLET) database \cite{cole1994isolet}:
\begin{itemize}
  \item $/ay/$  male speaker, CF = $1.5$\,kHz (data-set \texttt{fcmc0-A1}),
  \item $/es/$ female speaker, CF = $6.15$\,kHz (\texttt{fmb0-S1}).
\end{itemize}
%
Since the data is produced by simulations, we can utilize a large number of samples, $\cdot 10^5$, 
in our analysis.
%

\begin{figure}[h!]
\centering
\includegraphics[width=8 cm]{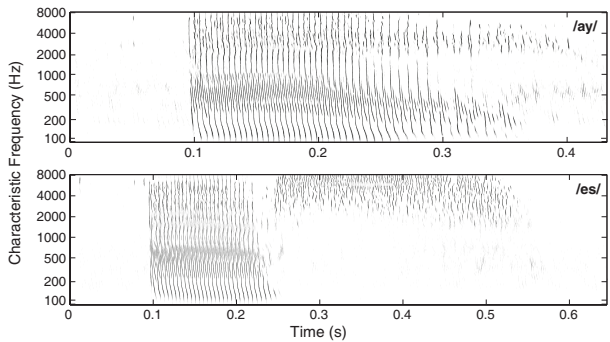}
\caption{Firing patterns of 40 globular bushy cells per characteristic frequency for $/ay/$ (top panel) and $/es/$(bottom panel). }
\label{fig:raster}
\end{figure}

Firing patterns (neurogram) of GBCs with CFs covering a part of the hearing range produced by the model are plotted in Fig. \ref{fig:raster}.
The neurogram exhibits is regular for the voiced parts of the speech sound (/$ay$/ and /$e$/), where the frikative /$s$/ has noisy components in the frequency range above about 2\,kHz.
The most striking difference of the neurogram compared to a 
spectrogram is its distinct temporal structure.
During the voiced parts of the speech sound (/$ay$/ and /$e$/), the firing patterns lock to the pitch frequency of the speaker and its higher harmonics with high fidelity.
During the frikative /$s$/, which has noise-like components in the frequency range above about 2\,kHz, the firing pattern was irregular.

\section{Entropy}
\label{sec:Entropy}

Entropy \cite{shannon2001mathematical} is a measure of the uncertainty associated with  a Random Variable (RV) $X$:
\ea{
H(X)& = -\sum_{x \in \Xcal} P[ X=x ] \log_2  P[ X=x ]
\label{eq:entropy definition}
}
where $\log_2$ is the logarithm to the base two and $\Xcal$ is the support of the RV $X$, i.e.  set of possible values of $X$.

In \cite{shannon2001mathematical}, Shannon defined entropy as a suitable measure of the uncertainty of a random source by following an axiomatic approach, that is by defining a series of desired properties for the measure be:

\begin{description}
  \item[Continuous in the probability distribution] Small variations in the probability distribution of $X$ correspond to small variation in $H(X)$
 \item[The uniform distribution maximizes entropy] This is the case where there is the most uncertainty on the random outcome
  \item[It should factorize] If a choice be broken down into two successive choices, the original $H$ should be the weighted sum
of the individual values of $H$
\end{description}

Shannon argued that the definition of entropy in \eqref{eq:entropy definition}, apart from scaling, is the only definition consistent with these axioms.
%
Other important properties make the definition of entropy in \eqref{eq:entropy definition} a particularly suitable measure of random predictability.
In particular:

\begin{description}
  \item [ Data processing inequality: ]
  $
  H(X) \geq H( f (X) ),
  $
  where $f$ is any function.
  That is entropy is reduced by any processing of the original RV.
  Entropy is preserved by injective transformations in which case the above inequality holds with equality.

  \item [ Boundedness: ]
  $
  0 \leq H(X) \leq \log_2 |\Xcal|
  $
The value of entropy is always bounded between the entropy of a degenerate distribution, a constant, and the uniform distribution over the support, $\log_2 |\Xcal|$.
\end{description}

%

It is also possible to define a conditional analogous to the entropy in \eqref{eq:entropy definition}:
\ea{
H (X|Y)
& = \sum_{x \in \Xcal, \ y \in \Ycal} P[ Y=y] H[ X | Y=y]
\label{eq:conditional entropy definition}
}
An important property of $H(X|Y)$ is the following: 
\begin{description}
  \item {\bf Conditioning reduces entropy:}
    $
    H(X|Y) \leq H(X)
    $, 
    where equality holds if and only if $Y$ is independent from $X$
\end{description}

With the definitions \eqref{eq:entropy definition} and \eqref{eq:conditional entropy definition} one can express the entropy of t $N$ RVs $\lcb X_1...X_N \rcb $ as a function of conditional entropies of the the RVs $X_i$:
\ea{
H \lb X_1...X_N \rb = \sum_{i=1}^N H\lb X_i  | X_1..X_{i-1} \rb.
\label{eq:chain rule H}
}

\section{A Time-Varying and Time-Dependent Entropy Estimate}
\label{sec:Entropy Estimation}

Strong \cite{strong1998entropy} proposed a method of evaluating the entropy by considering the sliding windows of size $T$.
We take here the same approach and consider the problem of estimating $H \lb W_1... W_N \rb $, the entropy of a sequence of $N$ sliding windows $W_i$  of size $T$.
By applying \eqref{eq:chain rule H} we obtain
\ea{
H\lb W_1...W_N \rb & = \sum_{i=1}^N H \lb W_i | W_1...W_{i-1} \rb
\label{eq:conditionnig word}
}

From the expression in \eqref{eq:conditionnig word} we now see the difficulty in estimating the entropy of the whole firing sequence.
A precise estimate would require us consider the conditional entropy of the current word given all the past words.
This is in general a very complicated task as it requires to consider the joint distribution between one word and all the past words.
Note that we cannot claim independence among words and drop the conditioning because the word at time $i$ shares common bins with all the $i-T$ previous words.
We can assume that the system has a finite memory, that is the 
past dependence is limited to 
a certain set of windows, that is
\ea{
 H\lb W_i  |   W_1...W_{i-1}\rb =  H\lb W_i  |  W_{i-M}...W_{i-1} \rb
\label{eq:finte window}
}
for some $M$.
One could apply the ``conditioning reduces entropy''  property of the conditional entropy to obtain an upper bound on the actual entropy estimate, that is
\eas{
 H\lb W_i  |  W_1...W_{i-1} \rb & \leq H\lb W_i  \rb \implies \\
 H \lb W_1...W_N \rb  & \leq \sum_{i=1}^N H \lb W_i \rb
\label{eq:bound conditional entorpy}
}

This latter bound on the entropy estimate justifies the 
direct method of \cite{strong1998entropy} as being an upper bound to the entropy estimate.

%
%

\subsection{A time dependent entropy estimate}

Given the loss in accuracy of the entropy estimate when ignoring the time dependency among words, we propose a time-dependent estimate
of the entropy which provides more accurate estimates than the direct method.
This also results in a time-varying estimate of the entropy of the words that provides interesting insights between the entropy of the word $W_i$ and the auditory
input at time $i$.

In particular

\begin{itemize}
  \item  we evaluate the entropy of each word $W_i$ across samples and obtain an estimate that we then correlate to the sensorial input.
  We show that 
  this time varying estimate provides significant insight on how the neurons code the sensory input.

  \item For each codeword $W_i$ we evaluate the conditional entropy given the past values of the codeword.
  By noting the variation of the entropy estimate as the length of the conditioning increases, 
  we investigate the time correlation among firings and, implicitly, estimate the memory of the random process.

  \item  We use the time-varying  and time-dependent estimate of the entropy to  obtain a better estimate of the entropy of the overall neuronal firing.

\end{itemize}

\section{Numerical Results}
\label{sec:Numerical Results}

\begin{figure}[h!]
\centering
\includegraphics[width=8 cm]{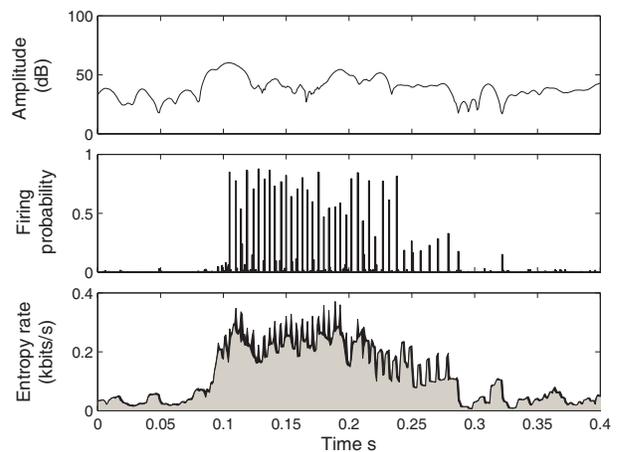}
\caption{Log-amplitude, average firing probability (1\,ms time bins),
 non-conditional (black) and conditional (grey) 
entropy estimate for $/ay/$ at 1.5kHz. Analysis window length: 10\,ms.}
\label{fig:logamp_a}
\end{figure}

In this section we present a set of numerical evaluation of the entropy estimate defined in Sec. \ref{sec:Entropy Estimation}. 
We begin by showing how the entropy estimate correlates to the log-amplitude of the frequency content at a given frequency.
In Fig. \ref{fig:logamp_a} we plot the short-term power spectrum 
at 1.5\,kHz for
the utterance $/ay/$ (top panel) together with the average firing probability (mid panel)  and the entropy estimate (lower panel).
For the entropy estimate we consider a window of length 10\,ms 
and we plot both the non-conditional version (black) of \eqref{eq:entropy definition}
and the conditional version (grey) 
of \eqref{eq:conditional entropy definition}, for which we use a conditioning over the past 20 windows.
From the plot it is clear that entropy estimates correlates to the frequency content at both low signal amplitudes, for the interval $0-0.1$\,s and $0.3-0.4$\,s,  as well as high signal amplitudes, for the interval $0.1-0.23$\,s.
The precise information on the temporal evolution of the frequency content is not provided by the firing probability.
This is an instantaneous measure of the neuronal activity that does not provide much information about the predictability of the response over time.
The conditional entropy estimate performs better than the non-conditional version as it is smoother in the interval $0.1-0.2$\,s and better follows the log-amplitude of the signal.
This is so because, in this interval, the probability of firing is very high and one obtains a better estimate by considering the correlation between the value of the current window and the past ones. 

\begin{figure}[h!]
\centering
\includegraphics[width=8 cm]{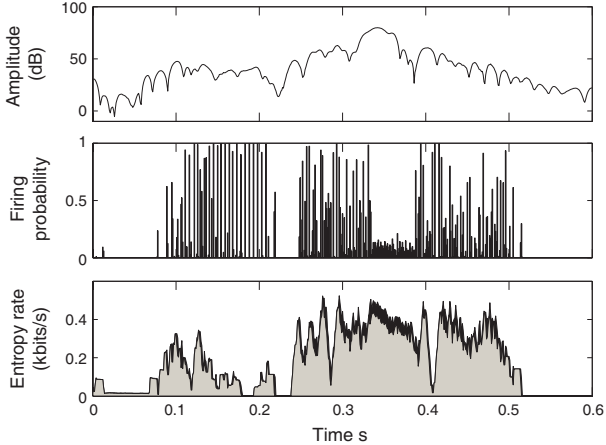}
\caption{Log-amplitude, average firing probability and entropy estimate for $/es/$ at 6.15\,kHz}
\label{fig:logamp_s}
\end{figure}

Fig. \ref{fig:logamp_s} is the analogous to Fig. \ref{fig:logamp_a} for the utterance of $/es/$.
As for Fig. \ref{fig:logamp_a} one observes a very precise time correlation between the entropy estimate and the log-amplitude of the frequency content.
For this case it is interesting to notice the behavior of the estimate in the interval $0.3-0.4$\,s. Even when 
the amplitude of the signal is large, the firing probability is very low,
despite of this the entropy estimate does not decay and preserve a good similarity to the amplitude content.
Another interesting detail is that the entropy decreases during the vowel part due to the regular firing pattern.
For the noise-like frikative /$s$/, the entropy rate is higher copmared to the vowel part.

\begin{figure}[h!]
\centering
\includegraphics[width=7 cm]{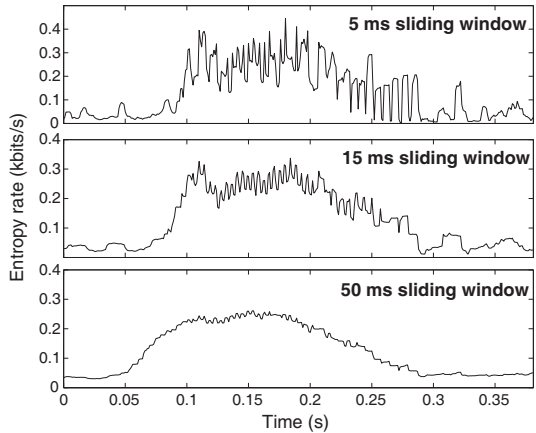}
\caption{The effect of the window length on the entropy estimate.}
\vspace{-.5 cm}
\label{fig:direct_method_entropy_diff_length}
\end{figure}

In Fig. \ref{fig:direct_method_entropy_diff_length} we study the impact of the window length on the entropy estimate for the $/ay/$ at 1.5 kHz.
For small windows the entropy estimate is very localized but suffers of great variations, as the predictability of a few bits can greatly increase the estimate.
For longer windows one obtains a smoother estimate but looses time resolution.

Finally we plot the cumulative difference between the entropy estimate and the conditional entropy estimate for different length of the conditioning.
For this plot we again consider the utterance  $/ay/$ at 1.5 kHz and a window length of 10\,ms. 
The error between $H(W_i)$  and $H(W_i|W_{i-L}...W_{i-1})$ as in \eqref{eq:bound conditional entorpy} is plotted for  $L=4,9$ and $20$\,ms. 
This difference indicates the refinement of the entropy estimate when using a time-dependent entropy estimate.
The impact of the conditioning also indicates the memory of the neuronal process.
A decrease in the entropy estimate for small values of $L$ is expected as we are accounting for the correlation of overlapping windows, that is windows that share the same bins.
A decrease of the conditional entropy for longer windows, instead, indicates a dependency among windows that do not share any bin and thus indicates a time dependency in the process.
%
This is indeed what we observe in Fig.   \ref{fig:direct_method_entropy_diff_length}: the conditioning for $L=20$\,ms indicates that in this case neuronal firings are correlated up to $20$\,ms. 

\begin{figure}[h!]
\centering
\includegraphics[width=7 cm]{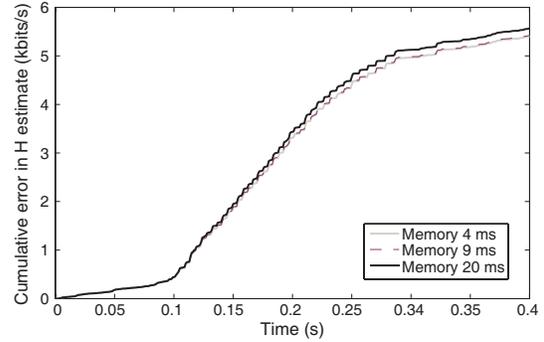}
\caption{Cummulative error between between $H(W_i)$  and $H(W_i|W_{i-L}...W_{i-1})$.}
\label{fig:cummulative_error}
\vspace{-.5 cm}
\end{figure}

\section{Conclusions}
\label{sec:Conclusions}

In this correspondence we introduce a time-dependent and time-varying entropy estimate as a measure of the predictability of 
neuronal responses.
We have for the first time analyzed the responses of globular bushy cells in the cochlear nucleus to voiced and unvoiced speech sounds.
These neurons extract temporal features of sound signals with high temporal fidelity, which is reflected in the extraordinary high entropy rates of their firing patterns.
%
Our method is 
particularly suitable 
for analyzing the neuronal response as 
it allows one to retain a high temporal resolution and yet consider long time dependencies in the signal.
This could not be attained with previous estimation approaches which would have either a precise temporal resolution, using short estimation windows, or account for time dependencies, using long estimation windows. 
%

\section{Acknowledgment}
The authors would like to thank Professor Gerhard Kramer for valuable discussions
and precious insights on the problem.
This work was funded within the BCCN Munich and  in the framework of an Alexander von Humboldt Professorship endowed by the German Federal Ministry of Education and Research (reference number 01GQ0441).

\bibliography{steBibNeuro}
\bibliographystyle{plain}

\end{document}